\documentclass[10pt,a4paper]{article}
\usepackage{amsmath}
\usepackage{amssymb}
\usepackage{amsfonts}
\usepackage{amstext}
\usepackage{amsbsy}
\usepackage[mathscr]{eucal}
\usepackage{graphicx}
\usepackage[all]{xy}
\usepackage{color}
\newcommand{\beq}{\begin{equation}}
\newcommand{\eeq}{\end{equation}}
\newcommand{\beqa}{\begin{eqnarray}}
\newcommand{\eeqa}{\end{eqnarray}}
\newcommand{\ket}[1]{| #1 \rangle}

\makeindex
\title{\Large\textbf{General multipartite entangled states and complex projective
variety}}

\author{\textit{ Hoshang Heydari}\\
        \small\textit{Institute of Quantum
Science, Nihon University,}\\
\small\textit{1-8 Kanda-Surugadai, Chiyoda-ku, Tokyo 101-8308, Japan
}}

\date{}
\begin{document}

\maketitle \thispagestyle{empty}

\begin{abstract}
We  discuss and investigate the geometrical structure  of general
multipartite states. In particular, we show that a geometrical
measure of entanglement for general multipartite states can be
constructed by the complex projective varieties defined by quadratic
polynomials.
\end{abstract}

%%%%%%%%%%%%%%%%%%%%%%%%%%%%%%%%%%%%%%%%%%%%%%%%%%
\section{Introduction} Quantum entanglement
 can  be
used as resource for performing some useful task such as quantum
cryptography and quantum teleportation which are classically
impossible. Moreover, entangled cluster states are very useful
building block of one-way quantum computer as a scheme for universal
quantum computation. In quantum mechanics the geometry of pure
quantum state is completely described by complex projective space
$\mathbf{CP}^{n}$ which is the set of all one-dimensional subspaces
of complex vector space $\mathbf{C}^{n+1}$. Moreover, the geometry
of bipartite and multipartite product states are described by a map
called Segre embedding of complex multi-projective spaces
\cite{Dorje99,Miyake,Hosh5,Hosh6,Beng}. In this paper we will
investigate geometrical structure of multipartite states. We will
also construct a measure of entanglement for general multipartite
states based on these complex projective varieties. In particular,
in the  section \ref{cpv} we will give a short introduction to the
complex affine algebraic and projective varieties. In the section
\ref{cpvms} we will define the Segre variety and some other complex
multi-projective varieties which enable us to construct a measure
entanglement for general multipartite states.
%%%%%%%%%%%%%%%%%%%%%%%%%%%%%%
But before that, denote a general, composite quantum system with $m$
subsystems as $\mathcal{Q}=\mathcal{Q}_{m}(N_{1},N_{2},\ldots,N_{m})
$ $=\mathcal{Q}_{1}\mathcal{Q}_{2}\cdots\mathcal{Q}_{m}$, with the
pure state $ \ket{\Psi}=\sum^{N_{1}}_{k_{1}=1}$
$\sum^{N_{2}}_{k_{2}=1} $
$\cdots\sum^{N_{m}}_{k_{m}=1}\alpha_{k_{1}k_{2}\ldots k_{m}}
\ket{k_{1}}\otimes\ket{k_{2}}\otimes\cdots\otimes \ket{k_{m}} $
defined on the Hilbert space $
\mathcal{H}_{\mathcal{Q}}=\mathcal{H}_{\mathcal{Q}_{1}}\otimes
\mathcal{H}_{\mathcal{Q}_{2}}\otimes\cdots\otimes\mathcal{H}_{\mathcal{Q}_{m}}
$, where the dimension of the $j$th Hilbert space is
$N_{j}=\dim(\mathcal{H}_{\mathcal{Q}_{j}})$.  In particular, we
denote a pure two-qubit state by $\mathcal{Q}^{p}_{2}(2,2)$, where
additional superscript $p$ indicates that we are considering pure
state of a quantum system. Next, let $\rho_{\mathcal{Q}}$ denote a
density operator acting on $\mathcal{H}_{\mathcal{Q}}$. The density
operator $\rho_{\mathcal{Q}}$ is said to be fully separable, which
we will denote by $\rho^{sep}_{\mathcal{Q}}$, with respect to the
Hilbert space decomposition, if it can  be written as $
\rho^{sep}_{\mathcal{Q}}=\sum^\mathrm{N}_{k=1}p_k
\bigotimes^m_{j=1}\rho^k_{\mathcal{Q}_{j}},~\sum^\mathrm{N}_{k=1}p_{k}=1
$
 for some positive integer $\mathrm{N}$, where $p_{k}$ are positive real
numbers and $\rho^k_{\mathcal{Q}_{j}}$ denotes a density operator on
Hilbert space $\mathcal{H}_{\mathcal{Q}_{j}}$. If
$\rho^{p}_{\mathcal{Q}}$ represents a pure state, then the quantum
system is fully separable if $\rho^{p}_{\mathcal{Q}}$ can be written
as
$\rho^{sep}_{\mathcal{Q}}=\bigotimes^m_{j=1}\rho_{\mathcal{Q}_{j}}$,
where $\rho_{\mathcal{Q}_{j}}$ is the density operator on
$\mathcal{H}_{\mathcal{Q}_{j}}$. If a state is not separable, then
it is said to be an entangled state.

\section{Complex projective variety} \label{cpv}
 In this short section we will give an introduction to the complex affine and projective varieties.
%%%%%%%%%%%%%%%%%%%%%%%%%%%%%%%%%
Let $\mathbf{C}$ be a complex algebraic field. Then an affine
$n$-space over $\mathbf{C}$ denoted $\mathbf{C}^{n}$ is the set of
all $n$-tuples of elements of $\mathbf{C}$. An element
$P\in\mathbf{C}^{n}$ is called a point of $\mathbf{C}^{n}$ and if
$P=(a_{1},a_{2},\ldots,a_{n})$ with $a_{j}\in\mathbf{C}$, then
$a_{j}$ is called the coordinates of $P$.
 Let $\mathbf{C}[z]=\mathbf{C}[z_{1},z_{2}, \ldots,z_{n}]$ denotes the polynomial
algebra in $n$  variables with complex coefficients. Then, given a
set of $q$ polynomials $\{g_{1},g_{2},\ldots,g_{q}\}$ with $g_{i}\in
\mathbf{C}[z]$, we define a complex affine variety as
\begin{eqnarray}
&&\mathcal{V}_{\mathbf{C}}(g_{1},g_{2},\ldots,g_{q})=\{P\in\mathbf{C}^{n}:
g_{i}(P)=0~\forall~1\leq i\leq q\}.
\end{eqnarray}
For example the space $\mathbf{C}^{n}$, the empty set and one-point
sets are trivial affine algebraic varieties given by
$\mathcal{V}_{\mathbf{C}}(0)=\mathbf{C}^{n}$,
$\mathcal{V}_{\mathbf{C}}(1)=\emptyset$, and
\begin{eqnarray}
&&\mathcal{V}_{\mathbf{C}}(z_{1}-a_{1},z_{2}-a_{2},\ldots,z_{n}-a_{n})=\{(a_{1},a_{2},\ldots,a_{n})\}.
\end{eqnarray}
Let $\mathcal{V}_{\mathbf{C}}$ be complex affine algebraic variety.
Then an ideal of $\mathbf{C}[z_{1},z_{2}, \ldots,z_{n}]$ is defined
by
\begin{eqnarray}
&&\mathcal{I}(\mathcal{V}_{\mathbf{C}})=\{g\in\mathbf{C}[z_{1},z_{2},
\ldots,z_{n}]: g(z)=0~\forall~z\in\mathcal{V}_{\mathbf{C}}\}.
\end{eqnarray}
 Note also that
 $\mathcal{V}_{\mathbf{C}}(\mathcal{I}(\mathcal{V}_{\mathbf{C}}))=\mathcal{V}_{\mathbf{C}}$.
 Moreover, we define a coordinate ring of an affine variety $\mathcal{V}_{\mathbf{C}}$
 by $C[\mathcal{V}_{\mathbf{C}}]=\mathbf{C}[z_{1},z_{2},
 \ldots,z_{n}]/\mathcal{I}(\mathcal{V}_{\mathbf{C}})$.
%%%%%%%%%%%%%%%%%%
 A complex projective space $\mathbf{CP}^{n}$ is
defined to be the set of lines through the origin in
$\mathbf{C}^{n+1}$, that is, $
\mathbf{CP}^{n}=\frac{\mathbf{C}^{n+1}-\{0\}}{
(x_{1},\ldots,x_{n+1})\sim(y_{1},\ldots,y_{n+1})},~\lambda\in
\mathbf{C}-0,~y_{i}=\lambda x_{i} ~\forall ~0\leq i\leq n+1 $. For
example $\mathbf{CP}^{1}=\mathbf{C}\cup\{\infty\}$,
$\mathbf{CP}^{2}=\mathbf{C}^{2}\cup\mathbf{CP}^{1}=\mathbf{C}^{2}\cup\mathbf{C}\cup\{\infty\}$
and in general we have
$\mathbf{CP}^{n}=\mathbf{C}^{n}\cup\mathbf{CP}^{n-1}$.

Given a set of homogeneous polynomials
$\{g_{1},g_{2},\ldots,g_{q}\}$ with $g_{i}\in C[z]$, we define a
complex projective variety as
\begin{equation}
\mathcal{V}(g_{1},\ldots,g_{q})=\{O\in\mathbf{CP}^{n}:
g_{i}(O)=0~\forall~1\leq i\leq q\},
\end{equation}
where $O=[a_{1},a_{2},\ldots,a_{n+1}]$ denotes the equivalent class
of point $\{\alpha_{1},\alpha_{2},\ldots,$
$\alpha_{n+1}\}\in\mathbf{C}^{n+1}$. We can view the affine complex
variety
$\mathcal{V}_{\mathbf{C}}(g_{1},g_{2},\ldots,g_{q})\subset\mathbf{C}^{n+1}$
as a complex cone over the complex projective variety
$\mathcal{V}(g_{1},g_{2},\ldots,g_{q})$. The ideal and coordinate
ring of complex projective variety can be defined in similar way as
in the case of complex algebraic affine variety by considering the
complex projective space and it's homogeneous coordinate.

\section{Multi-projective  variety and multipartite entanglement
measure} \label{cpvms}In this section, we will review the
construction of the Segre variety \cite{Hosh5}. Then, we will
discuss the construction of a measure of entanglement for general
multipartite states based on an extension of the definition of the
Segre variety. Let
$(\alpha^{i}_{1},\alpha^{i}_{2},\ldots,\alpha^{i}_{N_{i}})$  be
points defined on the complex projective space
$\mathbf{CP}^{N_{i}-1}$. Then the Segre map
\begin{equation}
\begin{array}{ccc}
  \mathcal{S}_{N_{1},\ldots,N_{m}}:\mathbf{CP}^{N_{1}-1}\times\mathbf{CP}^{N_{2}-1}
\times\cdots\times\mathbf{CP}^{N_{m}-1}&\longrightarrow&
\mathbf{CP}^{N_{1}N_{2}\cdots N_{m}-1}
\end{array}
\end{equation}
is defined by $
 ((\alpha^{1}_{1},\alpha^{1}_{2},\ldots,\alpha^{1}_{N_{1}}),\ldots,
 (\alpha^{m}_{1},\alpha^{m}_{2},\ldots,\alpha^{m}_{N_{m}}))  \longmapsto
 (\alpha^{1}_{i_{1}}\alpha^{2}_{i_{2}}\cdots\alpha^{m}_{i_{m}})$. Now, we let $\alpha_{i_{1}i_{2}\cdots i_{m}}$,$1\leq
i_{j}\leq N_{j}$ be a homogeneous coordinate-function on
$\mathbf{CP}^{N_{1}N_{2}\cdots N_{m}-1}$. Moreover, we consider the
general, pure composite quantum system
$\mathcal{Q}^{p}_{m}(N_{1},N_{2},\ldots,N_{m})$ and let $
\mathcal{A}=\left(\alpha_{i_{1}i_{2}\ldots i_{m}}\right)_{1\leq
i_{j}\leq N_{j}}, $ for all $j=1,2,\ldots,m$. $\mathcal{A}$ can be
realized as the following set $\{(i_{1},i_{2},\ldots,i_{m}):1\leq
i_{j}\leq N_{j},\forall~j\}$, in which each point
$(i_{1},i_{2},\ldots,i_{m})$ is assigned the value
$\alpha_{i_{1}i_{2}\ldots i_{m}}$. This realization of $\mathcal{A}$
is called an $m$-dimensional box-shape matrix of size $N_{1}\times
N_{2}\times\cdots\times N_{m}$, where we associate to each such
matrix a sub-ring
$\mathrm{S}_{\mathcal{A}}=\mathbf{C}[\mathcal{A}]\subset\mathrm{S}$,
where $\mathrm{S}$ is a commutative ring over the complex number
field. For each $j=1,2,\ldots,m$, a two-by-two minor about the
$j$-th coordinate of $\mathcal{A}$ is given by
\begin{eqnarray}\label{segreply1}
&&\alpha_{k_{1}k_{2}\ldots k_{m}}\alpha_{l_{1}l_{2}\ldots l_{m}} -
\alpha_{k_{1}k_{2}\ldots k_{j-1}l_{j}k_{j+1}\ldots
k_{m}}\alpha_{l_{1}l_{2} \ldots l_{j-1} k_{j}l_{j+1}\ldots l_{m}}\in
\mathrm{S}_{\mathcal{A}}.
\end{eqnarray}
Then the ideal $\mathcal{I}^{m}_{\mathcal{A}}$ of
$\mathrm{S}_{\mathcal{A}}$ is generated by this equation
 and
describes the separable states in $\mathbf{CP}^{N_{1}N_{2}\cdots
N_{m}-1}$. The image of the Segre embedding
$\mathrm{Im}(\mathcal{S}_{N_{1},N_{2},\ldots,N_{m}})$, which again
is an intersection of families of  hypersurfaces in
$\mathbf{CP}^{N_{1}N_{2}\cdots N_{m}-1}$, is called Segre variety
and it is given by
\begin{eqnarray}\label{eq: submeasure}
\mathrm{Im}(\mathcal{S}_{N_{1},N_{2},\ldots,N_{m}})&=&\bigcap_{\forall
j}\mathcal{V}(\alpha_{k_{1}k_{2}\ldots
k_{m}}\alpha_{l_{1}l_{2}\ldots l_{m}} \\\nonumber&&-
\alpha_{k_{1}k_{2}\ldots k_{j-1}l_{j}k_{j+1}\ldots
k_{m}}\alpha_{l_{1}l_{2} \ldots l_{j-1} k_{j}l_{j+1}\ldots l_{m}}).
\end{eqnarray}
We can also partition the Segre embedding as follows:
$$\xymatrix{ (\mathbf{P}^{N_{1}-1}\times\cdots\times\mathbf{P}^{N_{l}-1})
\times(\mathbf{P}^{N_{l}-1}\times\cdots\times\mathbf{P}^{N_{m}-1}
\ar[d]_{\mathcal{S}_{N_{1},N_{2},\ldots,N_{m}}})\ar[r]_{
\mathcal{S}_{N_{1},\ldots,N_{l}}\otimes
I}&\mathbf{P}^{\mathcal{M}_{1}}
\times(\mathbf{P}^{N_{l}-1}\times\cdots\times\mathbf{P}^{N_{m}-1})
\ar[d]_{I\otimes \mathcal{S}_{N_{l+1},\ldots,N_{m}}}\\
             \mathbf{P}^{N_{1}N_{2}\cdots N_{m}-1}&\mathbf{P}^{\mathcal{M}_{1}}\times\mathbf{P}^{\mathcal{M}_{2}}
             \ar[l]_{\mathcal{S}_{\mathcal{M}_{1},\mathcal{M}_{2}}}}$$
%%%%%%%%%%%%%%%%%%%%%%%%%%%%%%%%%%%%%%%%%%%%%%%%%%%%
where $\mathcal{M}_{1}=N_{1}N_{2}\ldots
N_{l}-1$,$\mathcal{M}_{2}=N_{l+1}N_{l+2}\ldots N_{m}-1$,
$(\mathcal{M}_{1}+1)(\mathcal{M}_{2}+1)=N_{1}N_{2}\ldots N_{m}$, and
$\mathbf{P}$ denotes the complex projective space. For the Segre
variety, which is represented by a completely decomposable tensors,
the above diagram commutate, that is
$\mathcal{S}_{N_{1},N_{2},\ldots,N_{m}}=(\mathcal{S}_{N_{1},\ldots,N_{l}}\otimes
I)\circ (I\otimes \mathcal{S}_{N_{l+1},\ldots,N_{m}})\circ
\mathcal{S}_{N_{1}N_{2}\ldots N_{l}-1,N_{l+1}N_{l+2}\ldots N_{m}-1}
$.
%%%%%%%%%%%%%%%
 Now, we define an entanglement
measure for a pure multipartite state as
\begin{eqnarray}\label{EntangSeg}\nonumber
\mathcal{E}(\mathcal{Q}^{p}_{m}(N_{1},\ldots,N_{m}))&=&(\mathcal{N}\sum_{
k_{j},l_{j}, j=1,2,\ldots,m}|\alpha_{k_{1}k_{2}\ldots
k_{m}}\alpha_{l_{1}l_{2}\ldots l_{m}}
\\&&-
\alpha_{k_{1}k_{2}\ldots k_{j-1}l_{j}k_{j+1}\ldots
k_{m}}\alpha_{l_{1}l_{2} \ldots l_{j-1}k_{j}l_{j+1}\ldots
l_{m}}|^{2})^{\frac{1}{2}},
\end{eqnarray}
where $\mathcal{N}$ is a normalization constant and
$j=1,2,\ldots,m$. This measure coincides with the generalized
concurrence for a general bipartite and 3-partite state, but for
reasons that we have explained in \cite{Hosh5}, it fails to quantify
the entanglement for $m\geq 4$, whereas it still provides the
condition of full separability. However, it is still possible to
define an entanglement measure for general multipartite states if we
modify equation (\ref{EntangSeg}) in such a way that it contains all
possible permutations of indices. In our recent paper \cite{Hosh6}
we have proposed a measure of entanglement for general pure
multipartite states as
\begin{eqnarray}\label{EntangSeg2}\nonumber
&&\mathcal{F}(\mathcal{Q}^{p}_{m}(N_{1},\ldots,N_{m}))
=(\mathcal{N}\sum_{\forall \sigma\in\text{Perm} (u)}\sum_{
k_{j},l_{j}, j=1,2,\ldots,m}\\&&|\alpha_{k_{1}k_{2}\ldots
k_{m}}\alpha_{l_{1}l_{2}\ldots l_{m}} -
\alpha_{\sigma(k_{1})\sigma(k_{2})\ldots\sigma(k_{m})}\alpha_{\sigma(l_{1})\sigma(l_{2})
\ldots\sigma(l_{m})}|^{2})^{\frac{1}{2}},
\end{eqnarray}
where $\sigma\in\text{Perm}(u)$ denotes all possible sets of
permutations of indices for which $k_{1}k_{2}\ldots k_{m}$ are
replace by $l_{1}l_{2}\ldots l_{m}$, and $u$ is the number of
indices to permute. By construction this measure of entanglement
vanishes on product states and it is also invariant under all
possible permutations of indices. Note that the first set of
permutations defines the Segre variety, but there are also
additional complex projective varieties embedded in
$\mathbf{CP}^{N_{1}N_{2}\cdots N_{m}-1}$ which are defined by other
sets of permutations of indices in equation (\ref{EntangSeg2}).
These varieties are defined by similar quadratic polynomials as
those used to define the Segre variety. For example  these varieties
 are defined by
\begin{eqnarray}\label{eq: submeasure}
\mathrm{T}_{\overline{N}}&=&\bigcap_{\forall \sigma\in\text{Perm}
(u), k_{j},l_{j},
j=1,2,\ldots,m}\mathcal{V}(\alpha_{k_{1},k_{2},\ldots,k_{m}}\alpha_{l_{1},l_{2},\ldots,l_{m}}\\\nonumber&&-
\alpha_{\sigma(k_{1})\sigma(k_{2})\ldots\sigma(k_{m})}\alpha_{\sigma(l_{1})\sigma(l_{2})
\ldots\sigma(l_{m})})
\end{eqnarray}
which  also include the Segre variety. As we have shown the Segre
variety is defined by completely decomposable tensor but these new
varieties are can be partially decomposable tensors. %%%%
 We can also
apply the same procedure as in the case of the concurrence
 to define a measure of entanglement for arbitrary multipartite
 states
\begin{eqnarray}\label{EntangSeg2}
\mathcal{F}(\mathcal{Q}_{m}(N_{1},\ldots,N_{m}))&=&\inf\sum_{i}p_{i}\mathcal{F}^{i}(\mathcal{Q}^{p}_{m}(N_{1},\ldots,N_{m}))
\\\nonumber&=&\inf\sum_{i}p_{i}(\mathcal{N}\sum_{\forall \sigma\in\text{Perm}
(u)}\sum_{ k_{j},l_{j}, j=1,2,\ldots,m}|\alpha^{i}_{k_{1}k_{2}\ldots
k_{m}}\alpha^{i}_{l_{1}l_{2}\ldots l_{m}}\\\nonumber&& -
\alpha^{i}_{\sigma(k_{1})\sigma(k_{2})\ldots\sigma(k_{m})}\alpha^{i}_{\sigma(l_{1})\sigma(l_{2})
\ldots\sigma(l_{m})}|^{2})^{\frac{1}{2}}.
\end{eqnarray}
where infimum  are taken over all pure decompositions of the density
matrix $\rho_{Q}$ acting on the Hilbert space
$\mathcal{H}_{\mathcal{Q}}$. However, this operating is not a easy
task to perform. Thus, we have shown that we can easily use advanced
mathematical tool of algebraic geometry such as complex
multi-projective varieties  to construct measure of entanglement for
general multipartite states.

\begin{flushleft}
\textbf{Acknowledgments:} The  author acknowledges the financial
support of the Japan Society for the Promotion of Science (JSPS).
\end{flushleft}

%%%%%%%%%%%%%%%%%%%%%%%%%%%%%%%%%%%%%%%%%%%%%%%%%%%%%%%%%%%%%%%%%%

\end{document}